\documentclass[final,5p,times,authoryear]{elsarticle}

\usepackage{amssymb}
\usepackage{amsmath}
\usepackage{float}
\usepackage{multicol}
\usepackage{pgfplots}
\usepackage[separate-uncertainty=true, multi-part-units=single]{siunitx}
\usepackage{dblfloatfix}
\usepackage[dvipsnames]{xcolor}
\usepackage{graphicx}
\usepackage{booktabs}
\usepackage{url}

\begin{document}

\begin{frontmatter}

\title{Combining quasi-static and high frequency experiments for the viscoelastic characterization of brain tissue} 

\author[1]{Laura Ruhland\corref{cor1}%
	}
\ead{laura.ruhland@fau.de}
\author[2]{Nina Reiter}
\author[2]{Silvia Budday}
\author[1]{Kai Willner}
\ead{kai.willner@fau.de}

\cortext[cor1]{Corresponding author}
\address[1]{Institute of Applied Mechanics, Friedrich-Alexander-Universität Erlangen-Nürnberg, Egerlandstr. 5, 91058, Erlangen, Germany}
\address[2]{Institute of Continuum Mechanics and Biomechanics, 		Friedrich-Alexander-Universität Erlangen-Nürnberg, Dr.-Mack-Str. 81, 90762, Fürth, Germany}

\begin{abstract}
Mechanical models of brain tissue are a beneficial tool to simulate neurosurgical interventions, disease progression, or brain development. However, the accuracy and predictive capacity of such a model relies on a precise experimental characterization of the tissue’s mechanical behavior. Such a characterization is yet limited by inconsistent or contradictory experimental responses reported in the literature, particularly when measurements are performed in different time or length scales. Although brain tissue has been extensively investigated in previous studies, the combination of experimental findings from different scales has received limited attention. In this study, we combine \textit{ex vivo} mechanical responses of porcine brain tissue obtained at different time scales in a mechanical model. We investigated the mechanical behavior of three different brain regions in the quasi-static domain with multi-modal large strain rheometer measurements and at high frequencies with magnetic resonance elastography (MRE). A comparative analysis of the mechanical parameters obtained from both experimental techniques demonstrated consistent regional variations in the viscoelastic behavior across the two domains. However, the mechanical behavior changes from a higher elasticity in the quasi-static and low frequency domain to a dominating viscosity at high frequencies. Based on the quasi-static and the high frequency behavior, we calibrated a fractional Kelvin-Voigt model and consequently unified the two responses in a single mechanical model to obtain a comprehensive characterization of the tissue's mechanical behavior.

\end{abstract}


\end{frontmatter}

\section{Introduction}
\label{Intro}
Mechanical modeling of the brain is a valuable tool that can assist in better predicting  the progression of Alzheimer's disease \citep{fornari_prion-like_2019, schafer_interplay_2019}, traumatic brain injuries \citep{li_influence_2011, griffiths_finite_2022} and neurosurgical simulations \citep{sase_gpu-accelerated_2015}. Moreover, mechanical models of brains can enhance our understanding about brain development \citep{bayly_mechanical_2014, budday_physical_2015}. The reliability of such a model largely depends on an accurate characterization of the tissue's mechanical behavior though experimental studies. Yet, results reported in the literature vary considerably and are often inconsistent or contradicting, making the choice of appropriate material parameters challenging. The primary reason for these discrepancies is the tissue's biphasic composition and softness, resulting in a high sensitivity of the tissue to changes in the time and length scale \citep{budday_fifty_2020}. A particular challenge is therefore the characterization of brain tissue with experimental techniques that vary in those scales. Nevertheless, for a comprehensive brain model, it is essential to describe the tissue over an extended time and length range, and consequently combine experimental responses obtained with different testing techniques.\\

The viscoelastic behavior of brain tissue has been studied with multiple experimental techniques over the past decades. A review of those techniques is provided by \citet{faber_tissue-scale_2022}. To study the frequency-dependent material behavior of soft tissue, magnetic resonance elastography (MRE) \citep{muthupillai_magnetic_1995} and oscillating rheometry \citep{vappou_magnetic_2007, forte_characterization_2017} are widely employed. The quasi-static response of brain is often investigated with indentation \citep{budday_mechanical_2015} or multi-modal mechanical testing \citep{budday_mechanical_2017, hinrichsen_inverse_2023, reiter_human_2025}. Despite the large number of studies characterizing brain tissue in a single time scale, there are only a few studies combining experimental findings from different scales. Most comparative studies combined experiments in the lower frequency domain like oscillating rheometry or dynamic mechanical analysis with MRE, which has been done for brain tissue by \citet{vappou_magnetic_2007}. Further studies combined the mentioned techniques for other soft tissues, such as liver \citep{klatt_viscoelastic_2010} and substitute materials \citep{okamoto_viscoelastic_2011,arunachalam_quantitative_2017, sauer_collagen_2019}. Only a limited number of studies combined both experiments in the quasi-static and frequency domain for a comprehensive characterization \citep{ weickenmeier_brain_2018, chen_comparative_2021, ruhland_mechanical_2025}. \\

To accurately capture the mechanical behavior of brain tissue, it is important to distinguish between different brain regions in the experimental characterization \citep{griffiths_importance_2023, faber_tissue-scale_2022}. The brain is composed of three parts, the cerebrum, the cerebellum and the brain stem. On a macroscopic level, the brain can be subdivided into gray and white matter. However, both gray and white matter are heterogeneous on a microstructural level and the brain consists of a large number of anatomical regions of gray matter, white matter or a combination of both. For the cerebrum's gray matter, a distinction can be made between the cortex, where neurons are organized in layers, and the inner gray matter regions like the basal ganglia and the thalamus, where neurons are distributed in a less structured way. White matter tissues vary in their degree of axon orientation, with tracts such as the corpus callosum exhibiting strongly aligned axons, while regions with multiple crossing tracts, such as the corona radiata, contain axons of multiple orientations \citep{reiter_insights_2021, faber_tissue-scale_2022, reiter_human_2025}. Due to its micro-and macro-structural similarities to human brain, porcine brain has emerged as a valuable animal model for brain research \citep{sauleau_pig_2009} and many studies recorded a similar mechanical behavior in porcine brain to human brain \citep{weickenmeier_brain_2018, macmanus_towards_2020} \\

In this study, we characterize three porcine brain regions \textit{ex vivo} with two different experimental techniques that vary in their time scales. In large-strain rheometer experiments, we measured the quasi-static response of the brain tissue under compression, tension and torsional shear. Based on the measured material response, we estimated the mechanical parameters by inversely calibrating a hyper-viscoelastic material model in an ABAQUS simulation of the experiments to the quasi-static response. With a tabletop MRE system, we examined the high-frequency response of the brain. To compare the mechanical behavior in both domains, we calculated the storage and loss modulus from the material responses of both experimental techniques. Based on the dynamic moduli over the whole frequency range, we calibrated a fractional viscoelastic model, unifying the mechanical behavior of the two experimental techniques. This study is therefore the first to combine experiments in the quasi-static and very high frequency domain in a single mechanical model for different brain regions. 

\section{Methods}
\label{MatMet}
\subsection{Tissue preparation}
\begin{figure}[H]
    \centering
    \includegraphics{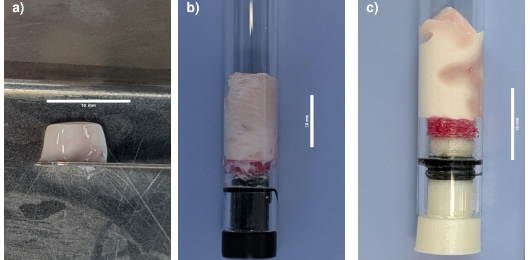}
    \caption{Exemplary samples of a) the rheometer experiment on brain, b) the tabletop MRE on the thalamus and c) the corona radiata with gray matter. The white bars have a length of \SI{10}{\milli\meter}}
    \label{fig:Sample}
\end{figure}

We examined the mechanical behavior of porcine brain provided by the local slaughterhouse Contifleisch GmbH and Unifleisch GmbH (Erlangen, Germany). Since the mechanical behavior of brain differs depending on the brain region \citep{budday_mechanical_2017}, we performed region specific tests for the brain tissue and extracted samples from three different brain regions, i.e., the corona radiata, the putamen, which is part of the basal ganglia, and the thalamus, from the porcine brains. For the rheometer experiments, we extracted cylindrical samples (Fig. \ref{fig:Sample} a)) with a diameter of \SI{8}{\milli\meter} and a height of \SI{4}{\milli\meter} from the tissue using a \SI{9}{\milli\meter} punch to consider the shape changes of the tissue during the cutting process \citep{reiter_human_2025}. To accurately set up the finite element model for the optimization, we measured the height and diameter of every sample after the punching process. The real sample dimensions were d\textsubscript{CR} = \SI{7.2 \pm 0.3}{\milli \meter} and h\textsubscript{CR} = \SI{4.6 \pm 0.3}{\milli \meter} for the corona radiata; d\textsubscript{P} = \SI{7.4 \pm 0.3}{\milli \meter} and  h\textsubscript{P} = \SI{4.5 \pm 0.3}{\milli \meter} for the putamen; and d\textsubscript{T} = \SI{7.3 \pm 0.4}{\milli \meter} and h\textsubscript{T} = \SI{4.3 \pm 0.4}{\milli \meter} for the thalamus. All rheometer measurements were conducted within \SIrange{1}{8}{\hour} post mortem. In the MRE the tissue was tested in glass tubes with an inner diameter of \SI{7}{\milli \meter} and a length of \SI{200}{\milli \meter}. We extracted cylindrical samples from the tissue with an \SI{8}{\milli \meter} punch, resulting in a sample diameter of \SI{7}{\milli \meter}. Exemplary brain samples are shown in Fig. \ref{fig:Sample} b) and c). The MRE measurements were performed between \SIrange{2}{3.5}{\hour} post-mortem. For both the rheometer and the MRE experiments, we used a new brain for the preparation of every sample. Before testing, the tissue was stored in the fridge at \SI{4}{\degreeCelsius} and moistened with Dulbecco's phosphate-buffered saline (DPBS) solution. We extracted the samples from the tissue directly before the measurement. 

\subsection{Rheometer}
\subsubsection{Experimental method}
To characterize the tissue in the quasi-static domain, we measured the material response under compression, tension, and torsional shear. For the experiments, we used a Discovery HR-3 and HR-30 rheometer (TA instruments, New Castle, Delaware, USA) with parallel plates and an upper geometry of \SI{8}{\milli\meter} diameter. To prepare the instrument, we fastened a circular piece of sandpaper with a diameter of \SI{9}{\milli\meter} to the upper and lower geometry using superglue (Pattex Ultra Gel) and calibrated the instrument with the applied sandpaper. Afterwards, we applied superglue to the sandpaper on both geometries and glued the sample to the upper geometry. Eventually, we lowered the upper geometry until the sample was connected to the bottom sandpaper. To avoid a compression of the sample, we ensured that the axial force did not exceed \SI{0}{\newton}. All measurements were performed in a DPBS bath at \SI{37}{\degreeCelsius} to mimic the in vivo conditions. The testing protocol included three cycles of compression and tension with $0.85$ and $1.15$ stretch and a strain rate of \SI{0.01}{\per\second}, followed by relaxation experiments in compression and tension with the same stretch but \SI{0.025}{\per\second} strain rate and \SI{300}{\second} holding time. Finally, we performed two cyclic torsional shear test with three loading cycles per experiment. In the first test, a shear strain of $0.15$ was applied with a loading rate of \SI{0.333}{\radian\per\second} and the second test was performed for a shear strain of $0.3$ and a loading rate of \SI{0.1667}{\radian\per\second}. In total, we measured \si{10} samples per brain region. 

\subsubsection{Vicoelastic modeling and inverse material parameter identification}
The material parameters of the tissue's quasi-static response were obtained by inversely calibrating a simulation of the tissue's deformations in ABAQUS to the measured stress-strain response. To capture the nonlinear and time-dependent material response of the tissue, a compressible hyper-viscoelastic constitutive model was employed. For the time-independent response, the strain energy function $\Psi$ is split into an isochoric and a volumetric part
\begin{equation}
	\Psi=\Psi_{iso}+\Psi_{vol}.
\end{equation}
A modified formulation of the Ogden model \citep{ogden_large_1972} was used for the isochoric part
\begin{equation}
	\Psi_{iso} = \sum_{i=1}^{2}\frac{2\mu_{i}}{\alpha_{i}^2}(\bar{\lambda}_{1}^{\alpha_{i}}+\bar{\lambda}_{2}^{\alpha_{i}}+\bar{\lambda}_{3}^{\alpha_{i}}-3) 
\end{equation}
with the shear moduli $\mu_{i}$, the nonlinearity coefficients $\alpha_{i}$ and the isochoric principle stretches $\bar{\lambda}_{i} = J^{-\frac{1}{3}}\lambda_{i}$ with the volume ratio $J$. This model has already been used in previous studies to model the mechanical behavior of brain tissue \citep{budday_mechanical_2017, hinrichsen_inverse_2023}. To capture the compression-tension asymmetry of the brain tissue, we choose two Ogden terms, in which the first term has a nonlinearity parameter $\alpha_1 > 0$ and the second term a nonlinearity parameter $\alpha_2 < 0$. The volumetric part of the strain energy function is defined as
\begin{equation}
	\Psi_{vol}= \frac{\kappa}{2}(J^{el}-1)^2,
\end{equation}
where $\kappa$ is the bulk modulus and $J^{el}$ is the elastic part of the volume ratio \citep{doll_development_2000}. With the initial shear modulus $\mu_{0} = \sum_{i=1}^{2}\mu_{i}$, which corresponds to the classical shear modulus in terms of linear elasticity, and the Poisson ratio $\nu$, the bulk modulus results from the relation 
\begin{equation}\label{equ:bulk}
	\kappa = \frac{2 \mu_{0} (1+\nu)}{3(1-2\nu)}.
\end{equation}   
Since the experimental data does not contain any information about volumetric deformation, the bulk modulus $\kappa$ is not treated as a free model parameter, but calculated with Eq. \ref{equ:bulk} and a fixed Poisson ratio $\nu$. \citet{hinrichsen_inverse_2023} investigated the influence of two different Poisson ratios $\nu = \{0.45,0.49\}$ on the identified material parameters for human brain tissue and found no significant difference in the fitting error. In this study, we therefore choose a Poisson ratio $\nu = 0.49$ for the simulation. 
The time-dependent material behavior was modeled with a two-term Prony series \citep{tschoegl_phenomenological_1989}, since previous studies demonstrated that two Prony terms are sufficient to represent the time response of brain tissue \citep{budday_mechanical_2015,chen_comparative_2021}. The two-term Prony series is equal to a generalized Maxwell model with two Maxwell units in parallel to an elastic spring \citep{tschoegl_phenomenological_1989}. A Prony relaxation function was applied to the shear moduli
\begin{equation}\label{equ:PronyShear}
	\mu^R_i(t) =  \mu_i \left( 1-\sum_{k=1}^{2}g_k \left(1-e^{-\frac{t}{\tau_k}}\right)\right)
\end{equation}
and the bulk modulus
\begin{equation}\label{equ:PronyBulk}
	\kappa^R(t) =  \kappa \left( 1-\sum_{k=1}^{2}k_k \left(1-e^{-\frac{t}{\tau_k}}\right)\right)
\end{equation}
with the shear and bulk Prony parameters $g_k$ and $k_k$ at the corresponding relaxation times $\tau_k$.\\

With the Fourier transform of Eq. \ref{equ:PronyShear}, the material response in the frequency domain can be represented in terms of the frequency dependent storage and loss modulus
\begin{equation}\label{eq:StorageLossModulus}
	\begin{split}
		G'(\omega) & = \mu_\infty+\mu_0\sum_{k=1}^{2}\frac{g_k\tau_k^2\omega^2}{1+\tau_k^2\omega^2} \\
		G''(\omega) & = \mu_0\sum_{k=1}^{2}\frac{g_k\tau_k\omega}{1+\tau_k^2\omega^2}
	\end{split}
\end{equation}
with the long term shear modulus $\mu_\infty = \mu_0 \left(1-\sum_{k=1}^{2}g_k\right)$.\\

The 10 free material parameters $\{\mu_1, \alpha_1, \mu_2, \alpha_2, g_1, g_2, k_1, k_2, \\ \tau_1, \tau_2\}$ were identified in an inverse material parameter identification process, using the Nelder-Mead simplex algorithm with bounded variables \texttt{fminsearbnd} \citep{john_derrico_fminseachbnd_2025} in MATLAB Release 2023b. The identification procedure minimized the sum over the residuum $R_i$ of each loading mode, i.e., cyclic compression-tension, compression relaxation, tension relaxation and cyclic torsion. The residuum for a single loading mode is defined as 
\begin{equation}\label{eq:res}
	R_i = \frac{\sqrt{\sum_{k=1}^{n}\vert x_k^{meas}-x_k^{calc} \vert^2}}{\sqrt{\sum_{k=1}^{n}\vert x_k^{meas} \vert^2}},
\end{equation} 
where $x^{meas}$ correspond to the measured and $x^{calc}$  to the calculated data vector. The residuum indicates the goodness of the calibration. To statistically evaluate the measured response, every measurement is fitted individually. However, to obtain an average response of the material, it is not accurate to take the average over those identified parameters, since this can lead to a false nonlinear behavior of the material \citep{hinrichsen_inverse_2023}. Therefore, \citet{hinrichsen_inverse_2023} suggested to fit the

\begin{figure}[H]
\centering
	\includegraphics{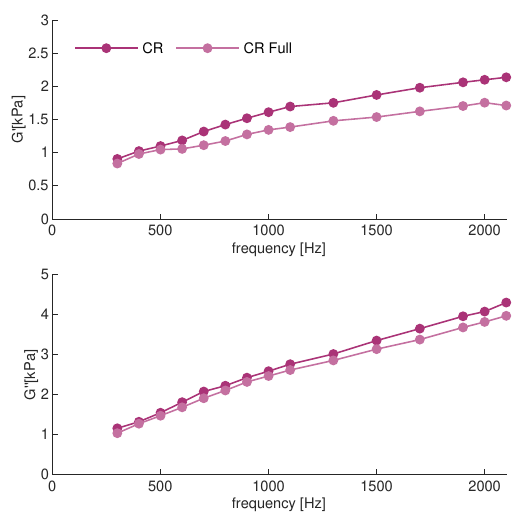}
	\caption{Comparison of the averaged storage and loss modulus for the corona radiata samples in which the full image was used for the data analysis (CR Full, $n = 7$) and in which gray matter parts were excluded from the image (CR, $n = 9$).}
	\label{fig:CRFull}
\end{figure}

average experimental curve instead, which we did in a second fitting procedure.

\subsection{Magnetic resonance elastography}
\subsubsection{Experimental method}
The material behavior in the frequency domain was studied with a tabletop MRE system. The experimental setup and the data acquisition procedure used in this study is equal to the MRE system and procedure described in \citep{braun_compact_2018}, which was already applied to study the frequency behavior of biological tissue \citep{braun_compact_2018} and hydrogels \citep{sauer_collagen_2019}. A \SI{0.5}{\tesla} permanent magnet MRI scanner (Pure Devices GmbH, Würzburg, Germany) was complemented with an external gradient amplifier (DC 600, Pure Devices GmbH, Würzburg, Germany) and a MRI system-controlled piezo-electric driver (PAHL 60/20, Piezosystem Jena GmbH, Jena Germany) to induce shear waves with a specific frequency into the sample. We measured the vibration response at 15 frequency points between \SI{300}{} and \SI{2100}{\hertz} with shear wave amplitudes between \SIrange{2.5}{6.6}{\micro\meter}. Wave images were acquired using a spin-echo MRE sequence and the image acquisition was carried out at four time steps per frequency. For a single frequency, the acquisition time was \SI{161 \pm 7.2}{\second}  for the corona radiata, \SI{181.2 \pm 20.1}{\second} for the putamen and \SI{177.7 \pm 11.9}{\second} for the thalamus. The repetition times for the different brain regions were TR\textsubscript{CR} = \SI{558.7\pm 86.4}{\milli\second}, TR\textsubscript{P} = \SI{639.2\pm 77.1}{\milli\second} and TR\textsubscript{T} = \SI{623.3\pm 46.9}{\milli\second}. Further imaging parameters were the same for all regions:  echo time TE = \num{27.8} - \SI{29.6}{\milli\second}, slice thickness: \SI{3}{\milli\meter}, field of view: \numproduct{9.6 x 9.6} \si{\milli\square\meter}, matrix size: \numproduct{64 x 64} and resolution: \numproduct{0.15 x 0.15 x 3} \si{\milli\cubic\meter}. All tests were conducted at \SI{37}{\degreeCelsius} and we measured 9 samples from the corona radiata and 5 from the putamen and the thalamus.

\subsubsection{Fractional viscoelastic modeling and model calibration}\label{sec:MREDataAnalysis}
The frequency dependent storage and loss moduli were determined by fitting an analytical solution of the shear wave based on the Bessel function to the wave image obtained for every frequency \citep{braun_compact_2018}. To determine the frequency-independent material parameters, a viscoelastic model was fitted to the storage and loss modulus. In this study we use a fractional Kelvin-Voigt model with two fractional elements in parallel to a spring to describe the material response at high frequencies. This model is a direct alternative to the generalized Maxwell model \citep{bonfanti_fractional_2020}, which was used to model the tissue's quasi-static response. The fractional Kelvin-Voigt model is defined as
\begin{equation}\label{eq:fracKV}
   G^*(\omega)=\mu_e +\sum_{i=1}^2 c_i(i\omega)^{\beta_i} 
\end{equation}
with the elastic shear modulus $\mu_e$ of the spring, the springpot constant $c_i$ and powerlaw exponent $0\leq \beta_i \leq 1$ of every springpot element. The springpot constant can be expressed in terms of the shear modulus $\mu_i$ and the viscosity $\eta_i$ 
\begin{equation}
    c_i = \mu_i^{1-\beta_i}\eta_i^{\beta_i}.
\end{equation}
Since $\mu_i$ and $\eta_i$ are linearly-dependent, we set $\eta = \SI{1}{\pascal \second}$ to obtain a single shear modulus for every springpot element \citep{klatt_viscoelastic_2010}. The spring in the fractional Kelvin-Voigt model defines the material behavior at \SI{0}{\hertz}. However, the measured frequency response does not contain any information about the response in the lower frequency regime. Therefore, we included the storage modulus obtained from the calibration of the rheometer measurements with Eq.\ref{eq:StorageLossModulus} as an additional parameter at \SI{0}{\hertz} to the MRE data. When two fractional elements are used in parallel, the fractional element with the lower exponent dominates the longer timescale, hence the lower frequencies, and the fractional element with the higher exponent dominates the shorter timescale, hence the higher frequencies \citep{bonfanti_fractional_2020}. Therefore, we include an additional condition , $\beta_1 < \beta_2$, in the identification process. For this calibration, the Nelder-Mead simplex algorithm with bounded variables and inequality constraints , \texttt{fminsearcon} \citep{john_derrico_fminseachbnd_2025}, was applied to identify the five free model parameters $\{\mu_e, \mu_1, \beta_1, \mu_2, \beta_2\}$. The identification process minimizes the residuum in Eq. \ref{eq:res} between the measured and the calculated data vector consisting of the data point obtained from the rheometer measurement and the MRE data, i.e., $x = \{ x_{rheo}, x_1, ... , x_n\}$. Similar to the rheometer measurements, we at first calibrate every measurement individually to statistically evaluate the data. Secondly, we calibrate the averaged measured data to obtain a single set of material parameters for the average frequency response. In case of the MRE measurements, we choose to calibrate the averaged measured data, since the individually identified material parameters are not normally distributed.  

\subsubsection{Corona radiata data processing}
 Some of the corona radiata samples contained gray matter on the edge of the cylindrical sample (Fig. \ref{fig:Sample} c)). To obtain the mechanical behavior of only the white matter, we selected a part from the MRE image that does not contain gray matter. Fig.\ref{fig:CRFull} shows the averaged modulus for the full data and the data in which the gray matter regions were excluded, revealing lower dynamic moduli for the full data. In previous MRE studies, a lower stiffness was reported for gray matter than for white matter \citep{weickenmeier_brain_2018}. Since for the MRE data analysis the shear wave is averaged over the radius of the sample, we assume that the gray matter part reduces the overall stiffness of the sample. We therefore use the corona radiata data without gray matter for the further analysis.

\subsection{Statistical analysis}
We performed group-wise comparisons between the three brain regions for the maximum and minimum stresses measured with the rheometer, the hyper-viscoelastic parameters and the fractional viscoelastic parameters. All statistical tests were done in MATLAB R2023b using the Statistics and Machine Learning Toolbox. The normal distribution was examined by applying a Shapiro-Wilk test using the \texttt{swtest} function \citep{bensaida_shapiro-wilk_2025}. Since not all data sets are normally distributed, non-parametric tests were employed to determine significant differences between the data groups. To determine the differences between the three brain regions, the Kruskal-Wallis test was applied. If significant differences were observed in the group-wise comparison, a pairwise post-hoc test was performed to identify the differences between the individual groups. As a post-hoc test, the Wilcoxon rank sum test with the Holm-Bonferroni correction method to control the family-wise error rate was performed, using the \texttt{fwer\_holmbonf} function \citep{martinez-cagigal_multiple_2025}. For all tests a \textit{p}-value lower than 0.05 was considered to be significant.

\begin{figure*}
    \centering
    \includegraphics[width=539pt]{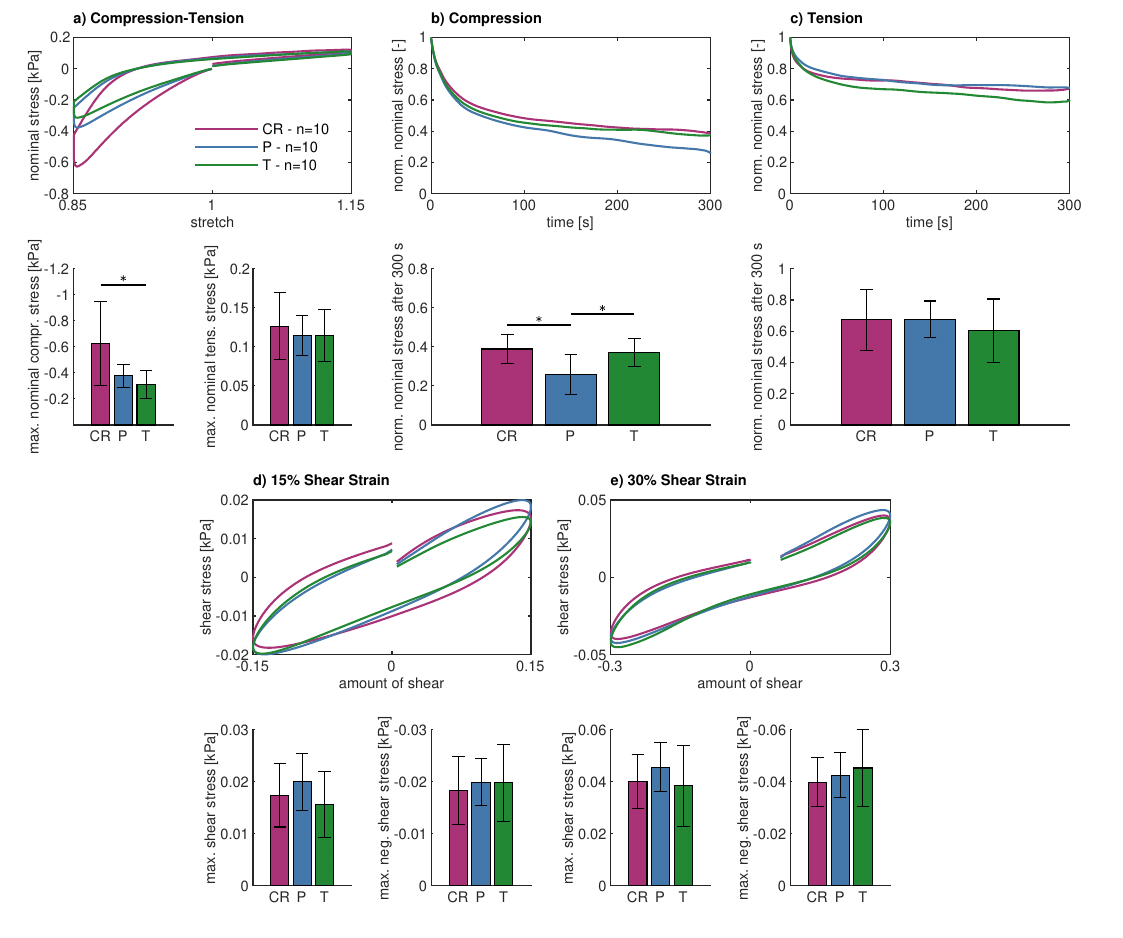}
    
    \caption{Comparison of the stress response obtained in multi-modality rheometer experiments for the corona radiata (CR), the putamen (P) and the thalamus (T). The averaged response under a) cyclic compression-tension, b) compression relaxation, c) tension relaxation, d) 15\% shear strain and e) 30\% shear strain with the corresponding maximum and minimum stresses under cyclic loading and the relaxed stress after \SI{300}{\second} in the relaxation test. Statistically significant differences between the groups are marked with '*' for $p < 0.05$.} \label{fig:rheo}
\end{figure*}

\begin{figure}[h]
\centering
	\includegraphics{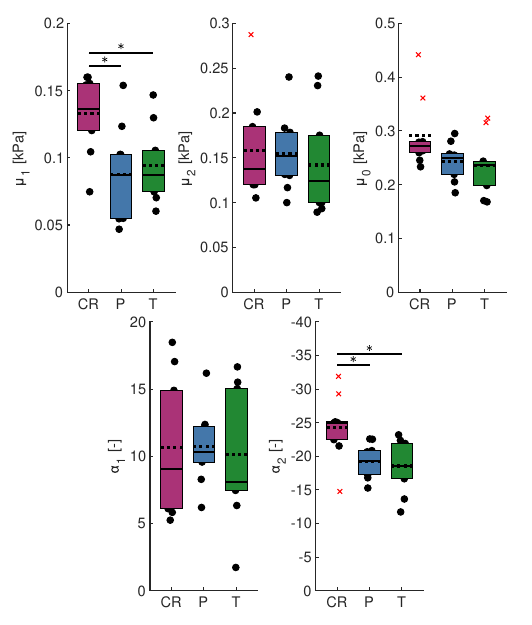}
	\caption{Comparison of hyperelastic parameters. Shear moduli $\mu_i$ and nonlinearity parameters $\alpha_i$ of the two-term Ogden model and the initial shear modulus $\mu_0$ for the corona radiata (CR), the putamen (P) and the thalamus (T). Solid lines donate the medians, dashed lines the mean values and outliers are marked in red. Statistically significant differences between the groups are marked with '*' for $p < 0.05$.}
	\label{fig:Ogden}
\end{figure}

\section{Results}\label{sec:results}
\subsection{Mechanical response in the quasi-static and low frequency domain}
Fig. \ref{fig:rheo} shows a comparison of the stress responses measured with the rheometer for the three brain regions the corona radiata, the putamen and the thalamus, as well as the maximum and minimum stresses during cyclic loading and the relaxed stress after \SI{300}{\second} of relaxation. Since under cyclic loading the regional trends for all three loading cycles were similar, we only plot the response of the first loading cycle for the comparison. The \textit{p}-values of the comparison are given in Tab. \ref{tab:pRheoMeas1} to \ref{tab:pRheoMeas5}. The differences between the three brain regions are the most pronounced under compression. The corona radiata behaves significantly stiffer than the thalamus and stiffer than the putamen at a significance level of \SI{0.1}{} (\textit{p} = 0.09). Moreover, the corona radiata has the most prominent compression-tension asymmetry, a characteristic behavior of brain tissue. In tension and torsional shear, all brain regions behave similarly. For stress relaxation in compression, the putamen relaxes significantly more than the corona radiata and the thalamus, whereas for tension relaxation there are no differences between the brain regions.\\

\begin{figure}[h]
\centering
	\includegraphics{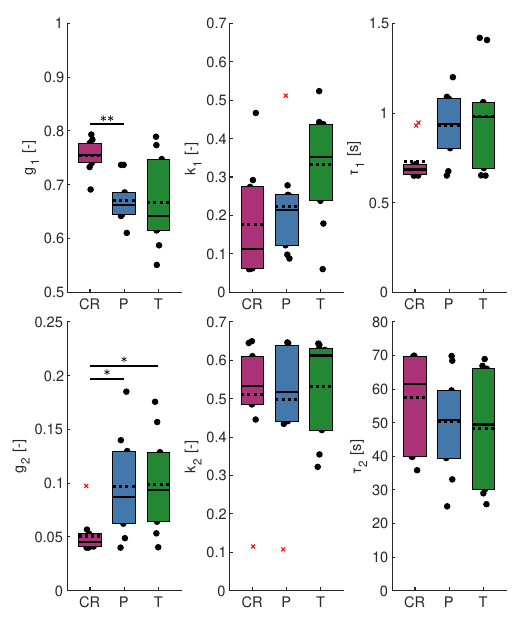}
	\caption{Comparison of the viscoelastic parameters. Prony parameters $g_i$ and $k_i$ and the relaxation times $\tau_i$ of the two-term Prony series for the corona radiata (CR), the putamen (P) and the thalamus (T). Solid lines donate the medians, dashed lines the mean values and outliers are marked in red. Statistically significant differences between the groups are marked with '*' for $p < 0.05$ and '**' for $p<0.01$.}
	\label{fig:Prony}
\end{figure}

The inversely determined mechanical parameters of the hyperelastic Ogden model are summarized in Fig. \ref{fig:Ogden} and the corresponding \textit{p}-values are listed in Tab. \ref{tab:pOgdenpara}. The corona radiata exhibits a significantly higher shear modulus $\mu_1$ of the first Ogden term than the putamen and the thalamus. For the shear modulus $\mu_0$, representing the overall stiffness of the tissue, the corona radiata again behaves stiffer than the other two regions at a significance lever of \SI{0.1}{} (\textit{p} = 0.09; \textit{p} = 0.08). The corona radiata region displays a higher negative nonlinearity parameter $\alpha_2$ than the other regions, corresponding to the stiffer compression response and the more pronounced compression-tension asymmetry of the corona radiata  observed in the measurements (Fig. \ref{fig:rheo}). \\

\begin{figure*}[h]
\centering
	\includegraphics{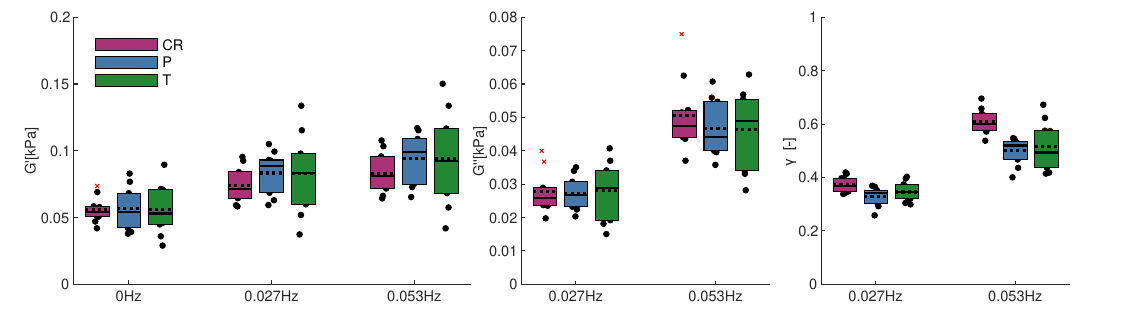}
	\caption{Comparison of the storage and loss modulus calculated from the Ogden-Prony parameters for the corona radiata (CR), the putamen (P) and the thalamus (T). The frequencies \SI{0.027}{\hertz} and \SI{0.053}{\hertz} correspond to the loading rates of the cyclic torsional shear tests. Solid lines donate the medians, dashed lines the mean values and outliers are marked in red.}
	\label{fig:rheoModu}
\end{figure*}

Fig. \ref{fig:Prony} contains a comparison of the identified viscoelastic material parameters of the Prony series and Tab. \ref{tab:pPronypara} the \textit{p}-values of the comparison. Significant differences are observed in the shear Prony parameters $g_1$ and $g_2$. At the shorter time scale, indicated by the smaller relaxation time $\tau_1$, the corona radiata exhibits a higher $g_1$ than the putamen at a significance level of \SI{0.05}{} and than the thalamus for a significance level of \SI{0.1}{} (\textit{p} = 0.05). Consequently, the corona radiata undergoes more relaxation at the shorter time scale. In the long term relaxation behavior, the putamen and the thalamus relax significantly more than the corona radiata, indicated by the higher $g_2$ values.\\

To compare the frequency responses of the brain regions, we calculated the storage and loss modulus at three frequency points \SI{0}{\hertz}, \SI{0.027}{\hertz} and \SI{0.053}{\hertz} with Eq. \ref{eq:StorageLossModulus}. The frequency points \SI{0.027}{\hertz} and \SI{0.053}{\hertz} correspond to the loading rates of the cyclic torsional shear tests. Fig. \ref{fig:rheoModu} shows that the frequency response is similar for all brain regions. The material parameters identified for the averaged measured material responses are summarized in Tab. \ref{tab:ogden} and Tab. \ref{tab:prony}. Based on those parameters, the storage moduli at \SI{0}{Hz}, which were used as an additional parameter for the model calibration of the MRE measurements, were calculated. The resulting storage moduli are listed in Tab. \ref{tab:storage0} .

\begin{table}[h]
	\centering
	\caption{Shear moduli and nonlinearity parameters of the two-term Ogden model for the corona radiata (CR), the putamen (P) and the thalamus (T).}
	\renewcommand\arraystretch{1.5}
	\begin{tabular}{lllll}
        \toprule
		& $\mu_1$ [kPa] & $\alpha_1$ [-] & $\mu_2$ [kPa]  & $\alpha_2$ [-]\\
		\midrule
		CR & 0.14 & 12.84 & 0.17 & -24.26 \\
		P & 0.16 & 6.07  & 0.14 & -20.19\\
		T & 0.12 & 9.21 & 0.1 & -18.35\\
		
	\end{tabular}
	\label{tab:ogden}
\end{table}

\begin{table}[h]
	\centering
	\caption{Prony parameters and relaxation times for the corona radiata (CR), the putamen (P) and the thalamus (T).}
	\renewcommand\arraystretch{1.5}
	\begin{tabular}{lllllll}
        \toprule
		& $g_1$ [-] & $k_1$ [-] & $\tau_1$ [s]  & $g_2$ [-] & $k_2$ [-] & $\tau_2$ [s]\\
		\midrule
		CR & 0.78 & 0.08 & 0.67 & 0.05 & 0.6 & 67.62\\
		P & 0.71 & 0.24  & 0.69 & 0.08 & 0.65 & 48.53\\
		T & 0.62 & 0.2 & 1.11 & 0.13 & 0.22 & 34.01\\
		
	\end{tabular}
	\label{tab:prony}
\end{table}

\begin{table}[h]
	\centering
	\caption{Storage modulus at \SI{0}{\hertz} for the corona radiata (CR), the putamen (P) and the thalamus (T) calculated based on the Ogden-Prony parameters.}
	\renewcommand\arraystretch{1.5}
	\begin{tabular}{lll}
        \toprule
		CR & P & T \\
		\midrule
		  \SI{0.057}{\kilo\pascal} & \SI{0.063}{\kilo\pascal} & \SI{0.052}{\kilo\pascal}\\
	\end{tabular}
	\label{tab:storage0}
\end{table}

\subsection{Mechanical response in the high frequency domain}
\begin{figure*}[h]
\centering
	\includegraphics{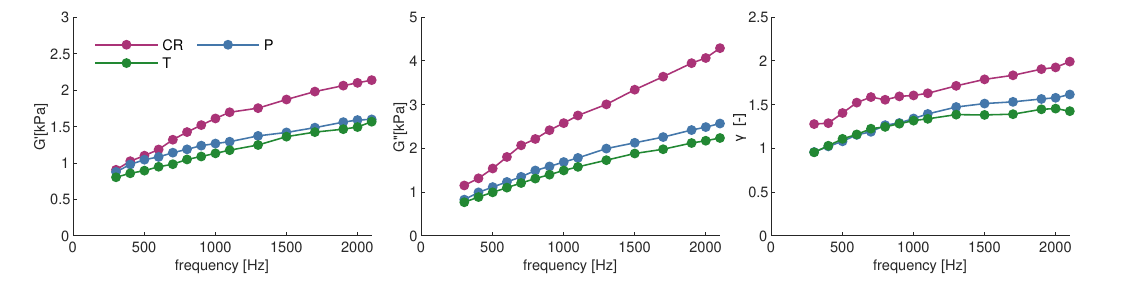}
	\caption{Mean storage modulus G', loss modulus G'' and loss tangent $\gamma$ for the corona radiata (CR, n = 9), the putamen (P, n = 5) and the thalamus (T, n = 5) measured with the tabletop MRE. The storage modulus describes the elastic and the loss modulus the viscous behavior of the material. The loss tangent is the ratio of the storage to the loss modulus $\gamma = G''/G'$ and indicates a more elastic material behavior for values $< 1$}
	\label{fig:MREModu}
\end{figure*}

Fig. \ref{fig:MREModu} presents the measured storage and loss moduli of the corona radiata, the putamen and the thalamus obtained from the tabletop MRE experiments as well as the loss tangents for frequencies from \SIrange{300}{2100}{\hertz}. For the corona radiata, the waves occasionally suffered from an insufficient wave stimulation at \SI{300}{\hertz}, resulting in nonphysical values for the storage and loss modulus. Therefore, we excluded some points of the \SI{300}{\hertz} data from the analysis, resulting in $n = 6$ data points for the corona radiata at \SI{300}{\hertz}. The storage and loss moduli of all regions demonstrate a continuous increase over the frequency range. We observe regional differences between the corona radiata and the other two brain regions. These differences become more pronounced with increasing frequency. At \SI{300}{\hertz}, the storage modulus was \SI{0.91 \pm 0.1}{\kilo\pascal} for the corona radiata, \SI{0.88 \pm 0.16}{\kilo\pascal} for the putamen, and \SI{0.8 \pm 0.07}{\kilo\pascal} for the thalamus. These values increased to \SI{2.14 \pm 0.27}{\kilo\pascal} for the corona radiata, \SI{1.6 \pm 0.31}{\kilo\pascal} for the putamen, and \SI{1.57 \pm 0.17}{\kilo\pascal} for the thalamus at \SI{2100}{\hertz}. Similarly, the loss modulus rose from  \SI{1.15 \pm 0.1}{\kilo\pascal} for the corona radiata,\SI{0.83 \pm 0.15}{\kilo\pascal} for the putamen, and \SI{0.77 \pm 0.11}{\kilo\pascal} for the thalamus at \SI{300}{\hertz} to \SI{4.29 \pm 0.9}{\kilo\pascal} for the corona radiata, \SI{2.57\pm 0.46}{\kilo\pascal} for the putamen, and \SI{2.24 \pm 0.3}{\kilo\pascal} for the thalamus at \SI{2100}{\hertz}. For all brain regions, the loss modulus exhibits a stronger frequency dependency than the storage modulus. Moreover, both the storage and loss modulus of the corona radiata are more frequency dependent than the dynamic moduli of the other two brain regions. This suggests that all regions get more viscous with increasing frequency and the corona radiata behaves more viscoelastic than the putamen and the thalamus.

\subsection{Viscoelastic parameter over a wide frequency range}
\begin{figure*}[h]
\centering
	\includegraphics{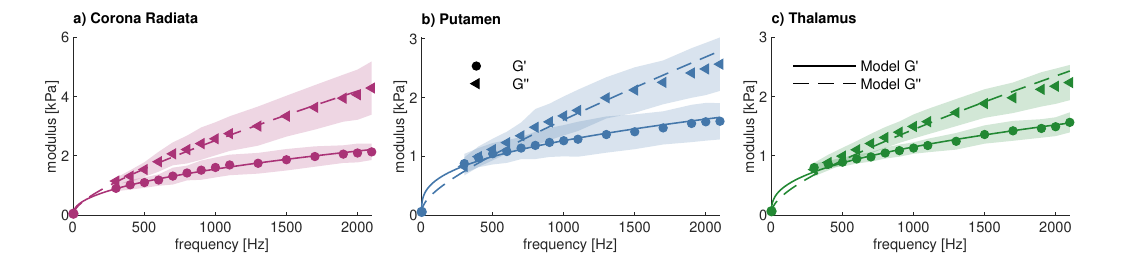}
	\caption{Fractional Kelvin-Voigt model calibrated for the averaged frequency data from \SIrange{0}{2100}{\hertz} for the corona radiata, the putamen and the thalamus. The storage modulus at \SI{0}{\hertz} was determined by rheometer experiments in the quasi-static domain and the high-frequency dynamic moduli (mean $\pm$ std) were measured with MRE.}
	\label{fig:compModelMeas}
\end{figure*}

\begin{figure*}[h]
\centering
	\includegraphics{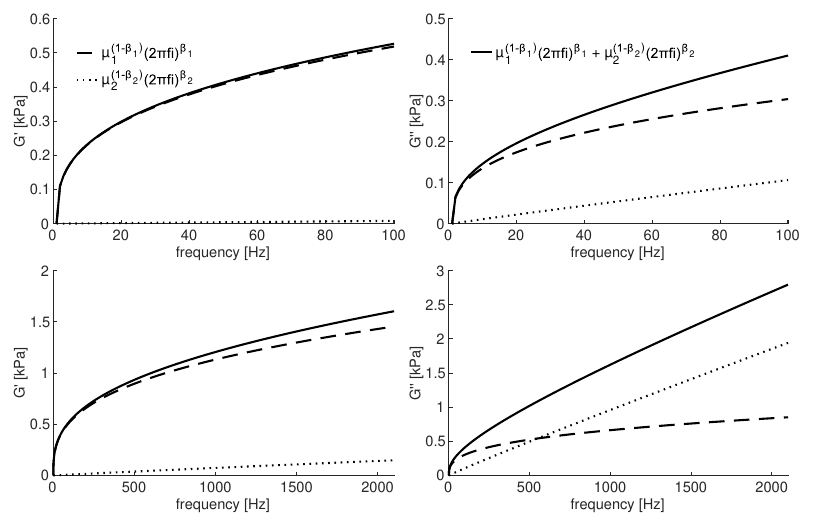}
    
	\caption{Contribution of the individual fractional elements to the overall storage (G') and loss modulus (G'') of the putamen. The low frequency response (\SIrange{0}{100}{\hertz}) is shown in the upper graph and the high-frequency response (\SIrange{0}{2100}{\hertz}) in the lower graph}\label{fig:InterprFracP}
\end{figure*}

To combine the high and low frequency response of brain tissue, we calibrated a fractional Kelvin-Voigt model with two fractional elements in parallel based on the dynamic moduli obtained from both the quasi-static rheometer measurements and the high-frequency MRE experiments. Fig. \ref{fig:compModelMeas} shows the model calibrated based on the averaged dynamic moduli and the dynamic moduli obtained from the measurements. Tab. \ref{tab:2fKV} contains the corresponding material parameters. The fractional model with two elements very well captures the response over the whole frequency range for all three brain regions. The first element combines a high shear modulus with a powerlaw exponent between \SI{0.3}{} and \SI{0.5}{}, while the second element consists of a very small shear modulus and a high powerlaw exponent between \SI{0.9}{} and \SI{1.0}{}. Typically, the fractional element with the lower exponent directs the lower frequency behavior, whereas the fractional element with the higher exponent controls the behavior at higher frequencies \citep{bonfanti_fractional_2020}. Fig. \ref{fig:InterprFracP} demonstrates the contribution of the individual fractional elements to the overall frequency response. Until approximately \SI{10}{\hertz}, both the storage and loss modulus are completely characterized by the first fractional element. For frequencies greater than \SI{10}{\hertz}, the storage modulus is still primarily shaped by the first fractional element. However, with increasing frequencies the contributing of the second fractional element to the storage modulus raises. On the contrary, the loss modulus is predominantly shaped by the second fractional element, which is moreover responsible for the accurate prediction of the higher frequency dependency of the loss modulus. For the interpretation of our material parameters, this means that the first fractional element mainly characterizes the storage modulus, thus the elastic response and the response at low frequencies, whereas the second fractional element  primarily characterizes the loss modulus, hence the viscous behavior and the high-frequency response.\\

\begin{figure*}[h]
\centering
	\includegraphics{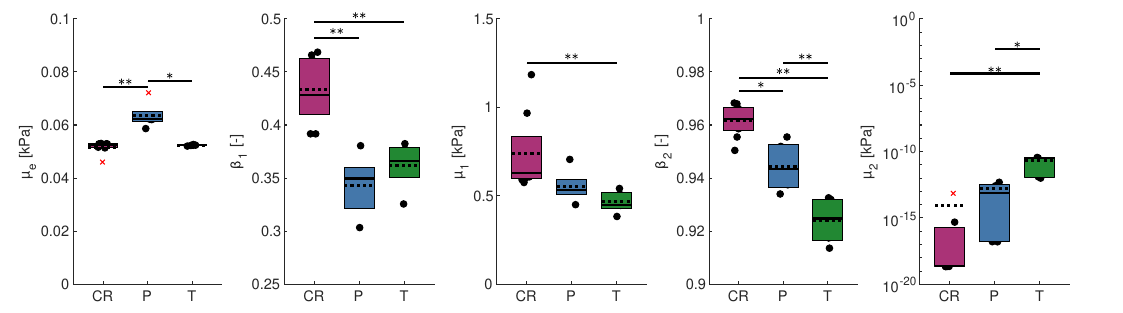}
	\caption{Comparison of the viscoelastic parameters in the frequency domain. A fractional Kelvin-Voigt model with two fractional elements (shear modulus $\mu_i$ and powerlaw exponent $\alpha_i$) in parallel to an elastic spring (elastic shear modulus $\mu_e$) was calibrated for the corona radiata (CR), the putamen (P) and the thalamus (T). The solid lines donate the medians, the dashed lines the mean values and outliers are marked in red. Statistically significant differences between the groups are marked with '*' for $p < 0.05$ and '**' for $p<0.01$.}
	\label{fig:MREPara}
\end{figure*}

Fig. \ref{fig:MREPara} compares the mechanical parameters identified for the individual measurements of the corona radiata,the putamen and the thalamus and Tab. \ref{tab:pMREpara} provides the corresponding \textit{p}-values of the comparison. The elastic shear moduli $\mu_e$ of the three brain regions correlate with the storage moduli obtained from the rheometer parameter identification in Tab. \ref{tab:storage0}, since those values were added to the data set for the model calibration, as described in Sec. \ref{sec:MREDataAnalysis}. For the first fractional element, regional differences mainly occur between the corona radiata and the other two brain regions. The corona radiata exhibits the highest first powerlaw exponent $\beta_1$ and a significantly higher shear modulus $\mu_1$ than the thalamus and the putamen for a significance level of \SI{0.1}{} (p = 0.058). For the second fractional element, we found considerable differences between all three brain regions in both the powerlaw exponent $\beta_2$ and the shear modulus $\mu_2$. The corona radiata has the largest powerlaw exponent, followed by the putamen and then the thalamus. In terms of the shear modulus, the thalamus demonstrates a higher value than the other two brain regions. Moreover, between the putamen and the corona radiata, differences very close to significance were found (p=0.0599), suggesting that the  putamen exhibits a larger shear modulus than the corona radiata. \\

\begin{table}[h]
	\centering
	\caption{Viscoelastic parameter for the fractional Kelvin-Voigt model with two fractional element parallel to a spring calibrated for the averaged frequency response of the corona radiata (CR), the putamen (P) and the thalamus (T).}
	\renewcommand\arraystretch{1.5}
	\begin{tabular}{lccccc}
        \toprule
		& $\mu_e$ [kPa] & $\beta_1$ [-] & $\mu_1$ [kPa]  & $\beta_2$ [-] & $\mu_2$ [kPa] \\
		\midrule
		CR & 0.053 &  0.463 & 0.689 & 0.964 & 2.23e-16 \\
 
		P & 0.063 &  0.337 & 0.591 & 0.951  & 1.21e-13 \\

		T & 0.052 &  0.370 & 0.464 & 0.933 & 2.13e-10 \\
	
	\end{tabular}
	\label{tab:2fKV}
\end{table}

 \section{Discussion}\label{sec:discu}
In this study, we mechanically characterized porcine brain tissue in the quasi-static and high-frequency domain and combined the frequency responses of both domains in a fractional Kelvin-Voigt model with two fractional elements in parallel. We investigated the material behavior of three distinct brain regions, i.e., the corona radiata, the putamen and the thalamus and found similar regional trends in both domains. In the quasi-static domain we obtained shear moduli of \SI{0.31}{\kilo\pascal} for the corona radiata, \SI{0.30}{\kilo\pascal} for the putamen and \SI{0.22}{\kilo\pascal} for the thalamus, indicating that the corona radiata is the stiffest brain region. Those moduli are within the range of shear moduli found in previous studies that investigated the quasi-static material response of porcine brain with indentation \citep{weickenmeier_brain_2018, gefen_are_2004}. Moreover, our measured stress-strain response (Fig. \ref{fig:rheo}) is consistent with a recent large-strain multi-modal study on porcine corona radiata and putamen \citep{reiter_mechanical_nodate}. Similar to our findings, the corona radiata was stiffer than the putamen under cyclical compression and tension. In the frequency domain, we observe as well that the corona radiata is the stiffest brain region with the highest overall shear modulus (Fig. \ref{fig:MREPara}). Previous MRE measurements conducted on porcine brain revealed that the cerebral white matter is stiffer than the cerebral gray matter, and more specifically, that the inner cerebral gray matter was stiffer than the outer gray matter \citep{weickenmeier_brain_2018}. These results support the regional differences we observe, as the corona radiata, which is part of the cerebral white matter, is behaving stiffer than the putamen, which belongs to the cerebral inner gray matter. The regional trends observed in this study differ from regional trends observed for human brain in earlier investigations. A stiffer material response was found in the human putamen and thalamus than in the corona radiata for both quasi-static large-strain multi-modal measurements \citep{hinrichsen_inverse_2023} and MRE \citep{hiscox_standardspace_2020}. However, this may be attributed to the differences in the species.  
For the viscoelastic material behavior, we found in the quasi-static domain a shorter relaxation time $\tau_1$, corresponding to a higher parameter $g_1$ and a longer relaxation time $\tau_2$, with a smaller parameter $g_2$, for all three brain regions (Fig. \ref{fig:Prony}). This aligns with viscoelastic parameters for porcine brain tissue in the literature \citep{gefen_are_2004} and indicates that brain tissue is more viscous at shorter time scales. Comparing the three brain regions, the corona radiata has the highest first relaxation parameter $g_1$, suggesting that the corona radiata is more viscous at shorter time scales than the putamen and the thalamus. Both observations in the quasi-static domain align with our findings in the frequency domain. The loss modulus increases with a faster rate than the storage modulus at high frequencies for all brain regions (Fig. \ref{fig:MREModu}). Thus, the material exhibits a more pronounced viscous behavior at high frequencies, which correspond to short time scales. Moreover, the corona radiata shows a higher loss modulus compared to the putamen and the thalamus at high frequencies, demonstrating a higher viscosity than the other two brain regions.  
To directly compare the mechanical responses obtained with the rheometer and the MRE, we calculated the low frequency storage and loss modulus from our identified material parameters in the quasi-static domain (Fig. \ref{fig:rheoModu}). Our low frequency dynamic moduli agree well with values recorded in the literature for porcine brain at \SI{0.1}{\hertz} \citep{vappou_magnetic_2007}. At high frequencies, our moduli are consistent as well with values in porcine brain stem for up to \SI{1500}{\hertz} \citep{braun_compact_2018} and in porcine the corona radiata for frequencies from \SIrange{80}{140}{\hertz} \citep{vappou_magnetic_2007}, while moduli remarkably larger than ours were found for porcine corpus callosum between \SI{200}{\hertz} and \SI{300}{\hertz} \citep{guertler_mechanical_2018} and porcine thalamus at \SI{80}{\hertz} \citep{weickenmeier_brain_2018}. In the latter study, the tissue was examined using a large-scale MRI scanner that is suitable for \textit{in situ} or \textit{in vivo} human studies. This differs from our study and the the investigations by \citet{vappou_magnetic_2007} and \citep{braun_compact_2018}, in which a compact tabletop MRI scanner, solely suitable for smaller brain sample, was used to characterize the tissue. There are two major differences between the large-scale and the tabletop MRE devices. Firstly, tabletop MRE scanners are operating on a lower field strength, usually \SI{0.5}{\tesla}, compared to the large-scale devices, where the field strange ranges form \SIrange{3}{7}{\tesla}. However, previous studies showed that mechanical parameters obtained from MRE are independent of the device's field strength \citep{ipek-ugay_tabletop_2015,zampini_measuring_2021}. Secondly, in the large scale devices the brain tissue can be examined \textit{in situ}, thus the structure of the brain is supported by the sculp and tested in the cerebral fluid, which may be a cause for the mechanical differences measured \textit{in situ} compared to our study. Comparing the dynamic moduli obtained from the quasi-static and high-frequency measurements, we observe a constant increase over the whole frequency range. However, the major differences between the low and the high frequency moduli is that at low frequencies, brain tissue exhibits a higher storage than loss modulus, whereas at high frequencies, the loss modulus exceeds the storage modulus. This difference can be further illustrated by the loss tangent, which is the ratio of the storage to the loss modulus and indicates the viscous dissipation of the material. For a loss tangent smaller than one, the material behaves more elastic, whereas a loss tangent larger than one indicates a more viscous material behavior. For brain tissue, we observe loss tangents below one for the low frequencies (Fig. \ref{fig:rheoModu}) and for high frequencies, the loss tangent is mostly larger than one (Fig. \ref{fig:MREModu}). This implies that the material behavior changes from an elasticity-dominated behavior at low frequencies to a viscosity-dominated behavior at high frequencies. To model this material behavior, we selected the fractional Kelvin-Voigt model with two fractional elements in parallel, since this configuration can model the change from a higher storage to a higher loss modulus with increasing frequency \citep{bonfanti_fractional_2020}. For the model calibration, we combined the quasi-static storage modulus from the rheometer measurements with the dynamic moduli obtained from the tabletop MRE. Although the rheometer measurements were conducted for large strains, resulting in a nonlinear deformation, the storage modulus only contains the elastic response of the material, since the storage modulus is related to the initial shear modulus $\mu_0$, which is the classical shear modulus in terms of the theory of linear elasticity. Consequently, the calibrated model combines the elastic response measured in both domains. A high agreement was achieved between the experimental data and the modeled data over the entire frequency range for all three brain regions (Fig.\ref{fig:compModelMeas}). The fractional Kelvin-Voigt model is therefore capable of modeling the wide frequency mechanical response of the brain and unites the quasi-static rheometer measurements with high frequency MRE.

\section{Conclusion}\label{sec:concl}
With this study, we aim to address the issue of inconsistent material response of brain tissue when experiments are conducted in different time scales. Therefore, we investigated the mechanical behavior of porcine brain in the quasi-static domain and the high-frequency domain. We observed consistent regional variations between the corona radiata, the putamen and the thalamus in both domains. However, a comparative study of the low and high frequency dynamic moduli revealed that the viscoelastic behavior of brain tissue changes from a dominating elasticity in the quasi-static domain and at low frequencies to a dominating viscosity at high frequencies. To combine the mechanical response in both domains, we calibrated a fractional Kelvin-Voigt model based on the quasi-static and high frequency data. The high agreement between the measured and the modeled response demonstrates that the fractional Kelvin-Voigt model accurately captures the mechanical behavior over a wide frequency range. Consequently, we demonstrated that the mechanical responses of brain tissue in different time scales can be unified with the fractional Kelvin-Voigt model, providing a comprehensive characterization of brain tissue over a wide frequency range.

\section*{Acknowledgments}
This work was funded by the German Research Foundation (DFG) project 460333672 CRC1540 EBM.

\bibliographystyle{elsarticle-harv} 
\bibliography{references}
\appendix
\section{Result of the statistical analysis}\label{app1}

See Tab. \ref{tab:pRheoMeas1} to \ref{tab:pMREpara}. 
\begin{table}[h]
	\centering
	\caption{Kruskal-Wallis \textit{p}-value for rheometer test in compression and tension}
	\begin{tabular}{cccc}
    \toprule
    \multicolumn{4}{l}{\textbf{max. nom. compr. stress}}\\
    \toprule
    & CR & P & T \\
    CR & - & 0.09 & 0.03 \\
    P &  & - & 0.14 \\
    T &  &  & - \\
    \midrule
    \multicolumn{4}{l}{\textbf{max. nom. tens. stress}}\\
    \toprule
    & CR & P & T \\
    CR & - & 1 & 1 \\
    P &  & - & 1 \\
    T &  &  & - \\
    \midrule
    \multicolumn{4}{l}{\textbf{norm. nom. stress after }}\\
    \multicolumn{4}{l}{\textbf{\SI{300}{s} of compression}}\\
    \toprule
    & CR & P & T \\
    CR & - & 0.03 & 0.57 \\
    P &  & - & 0.02 \\
    T &  &  & - \\
    \midrule
    \multicolumn{4}{l}{\textbf{norm. nom. stress after}}\\
    \multicolumn{4}{l}{\textbf{ \SI{300}{s} of tension}}\\
    \toprule
    & CR & P & T \\
    CR & - & 1 & 1\\
    P &  & - & 1 \\
    T &  &  & - \\
	\end{tabular}
	\label{tab:pRheoMeas1}
\end{table}

\begin{table}[h]
	\centering
	\caption{Kruskal-Wallis \textit{p}-value for rheometer test under torsional}
	\begin{tabular}{cccc}
    \toprule
    \multicolumn{4}{l}{\textbf{max. shear stress }}\\
    \multicolumn{4}{l}{\textbf{at 15\% shear strain}}\\
    \toprule
                & CR & P & T \\
                CR & - & 1 & 1 \\
                P &  & - & 1 \\
                T &  &  & - \\
    \midrule 
    \multicolumn{4}{l}{\textbf{max. neg. shear stress }}\\
    \multicolumn{4}{l}{\textbf{at 15\% shear strain}}\\
    \toprule
                    & CR & P & T \\
                CR & - & 1 & 1 \\
                P &  & - & 1 \\
                T &  &  & - \\
    \midrule
    \multicolumn{4}{l}{\textbf{max. shear stress }}\\
    \multicolumn{4}{l}{\textbf{at 30\% shear strain}}\\
    \toprule
                & CR & P & T \\
                CR & - & 1 & 1 \\
                P &  & - & 1 \\
                T &  &  & - \\
    \midrule 
    \multicolumn{4}{l}{\textbf{max. neg. shear stress}}\\
    \multicolumn{4}{l}{\textbf{at 30\% shear strain}}\\
    \toprule
                    & CR & P & T \\
                CR & - & 1 & 1 \\
                P &  & - & 1 \\
                T &  &  & - \\

	\end{tabular}
	\label{tab:pRheoMeas5}
\end{table}

\begin{table}[h]
	\centering
	\caption{Kruskal-Wallis \textit{p}-value for the Ogden parameters}
	\begin{tabular}{lcccc}
        \toprule
         $\mu_1$&  & CR & P & T \\
		        &  CR & - & 0.02 & 0.03 \\
                & P &  & - & 0.57 \\
                & T &  &  & - \\
        \midrule
         $\alpha_1$&  & CR & P & T \\
		        &  CR & - & 1 & 1\\
                & P &  & - & 01 \\
                & T &  &  & - \\
        \midrule
         $\mu_2$&  & CR & P & T \\
		        &  CR & - & 1 & 1\\
                & P &  & - & 1 \\
                & T &  &  & - \\
        \midrule
         $\alpha_2$&  & CR & P & T \\
		        &  CR & - & 0.01 & 0.01 \\
                & P &  & - & 0.85 \\
                & T &  &  & - \\
        \midrule
         $\mu_0$&  & CR & P & T \\
		        &  CR & - & 0.09 & 0.08 \\
                & P &  & - & 0.52 \\
                & T &  &  & - \\
        
	\end{tabular}
	\label{tab:pOgdenpara}
\end{table}

\begin{table}[h]
	\centering
	\caption{Kruskal-Wallis \textit{p}-value for the Prony parameters}
	\begin{tabular}{lcccc}
        \toprule
         $g_1$&  & CR & P & T \\
		        &  CR & - & 1.70e-3 & 0.05 \\
                & P &  & - & 0.79 \\
                & T &  &  & - \\
        \midrule
         $k_1$&  & CR & P & T \\
		        &  CR & - & 1 & 1 \\
                & P &  & - & 1 \\
                & T &  &  & - \\               
        \midrule
         $\tau_1$&  & CR & P & T \\
		        &  CR & - & 0.06 & 0.13 \\
                & P &  & - & 0.91 \\
                & T &  &  & - \\
        \midrule          
         $g_2$&  & CR & P & T \\
		        &  CR & - & 0.03 & 0.02 \\
                & P &  & - & 1 \\
                & T &  &  & - \\
        \midrule
         $k_2$&  & CR & P & T \\
		        &  CR & - & 1 & 1\\
                & P &  & - & 1 \\
                & T &  &  & - \\
        \midrule
         $\tau_2$&  & CR & P & T \\
		        &  CR & - & 1 & 1 \\
                & P &  & - & 1 \\
                & T &  &  & - \\
    
	\end{tabular}
	\label{tab:pPronypara}
\end{table}

\begin{table}[h]
	\centering
	\caption{Kruskal-Wallis \textit{p}-value for the 2-fKV parameters}
	\begin{tabular}{lcccc}
        \toprule
         $\mu_0$&  & CR & P & T \\
		        &  CR & - & 3.00e-3 & 0.80 \\
                & P &  & - & 0.02 \\
                & T &  &  & - \\
        \midrule
         $\beta_1$&  & CR & P & T \\
		        &  CR & - & 3.00e-3 & 2.00e-3 \\
                & P &  & - & 0.31 \\
                & T &  &  & - \\
        \midrule
         $\mu_1$&  & CR & P & T\\
		        &  CR & - & 0.06 & 3.00e-3 \\
                & P &  & - & 0.10 \\
                & T &  &  & - \\
        \midrule
         $\beta_2$&  & CR & P & T \\
		        &  CR & - & 0.01 & 3.00e-3 \\
                & P &  & - & 7.94e-3 \\
                & T &  &  & - \\
        \midrule
         $\mu_2$&  & CR & P & T \\
		        &  CR & - & 0.06 & 3.0e-3 \\
                & P &  & - & 0.02 \\
                & T &  &  & - \\

	\end{tabular}
	\label{tab:pMREpara}
\end{table}

\end{document}